\documentclass[%
superscriptaddress,
 amsmath,amssymb,
prx,
floatfix,
twocolumn
]{revtex4-2}

\usepackage{graphicx}
\usepackage{dcolumn}
\usepackage{bm}
\usepackage{booktabs}
\usepackage{multirow}
\usepackage{braket}
\usepackage{xcolor}
\usepackage{makecell}
\usepackage[colorlinks]{hyperref}
\usepackage{pifont}
\usepackage{placeins}
\usepackage{lineno}

\hypersetup{
citecolor={blue},
urlcolor={blue}
}

\newcommand{\cmark}{\ding{51}}%
\newcommand{\xmark}{\ding{55}}%

\begin{document}


\preprint{APS/123-QED}

\title{Symmetry Rules for Cavity Materials Engineering with Linearly Polarized Vacuum Fields
}

\author{Jingkai Quan}
\email{jingkai.quan@mpsd.mpg.de}
\affiliation{%
Max Planck Institute for the Structure and Dynamics of Matter,
Luruper Chaussee 149, 22761, Hamburg, Germany
}
\affiliation{The Hamburg Centre for Ultrafast Imaging, Hamburg, Germany}

\author{Chongxiao Fan}%
\email{chongxiao.fan@mpsd.mpg.de}
\affiliation{%
Max Planck Institute for the Structure and Dynamics of Matter,
Luruper Chaussee 149, 22761, Hamburg, Germany
}
\affiliation{Institute for Theory of Statistical Physics, RWTH Aachen University, and JARA Fundamentals of Future Information Technology, 52062 Aachen, Germany}
\author{Benshu Fan}
\email{benshu.fan@mpsd.mpg.de}
\affiliation{%
Max Planck Institute for the Structure and Dynamics of Matter,
Luruper Chaussee 149, 22761, Hamburg, Germany
}%
\author{I-Te Lu}
\affiliation{%
Max Planck Institute for the Structure and Dynamics of Matter,
Luruper Chaussee 149, 22761, Hamburg, Germany
}%
\affiliation{Department of Electrophysics, National Yang Ming Chiao Tung University, Hsinchu 300093, Taiwan}

\author{Dante M. Kennes}
\affiliation{Institute for Theory of Statistical Physics, RWTH Aachen University, and JARA Fundamentals of Future Information Technology, 52062 Aachen, Germany}
\affiliation{%
Max Planck Institute for the Structure and Dynamics of Matter,
Luruper Chaussee 149, 22761, Hamburg, Germany
}%
\author{Angel Rubio}
 \email{angel.rubio@mpsd.mpg.de}
\affiliation{%
Max Planck Institute for the Structure and Dynamics of Matter,
Luruper Chaussee 149, 22761, Hamburg, Germany
}%
\affiliation{Initiative for Computational Catalysis, The Flatiron Institute, Simons Foundation, New York City, NY 10010, United States of America}

\date{\today}

\begin{abstract}
Cavity materials engineering, aiming to manipulate material properties by coupling to vacuum fluctuations inside a cavity, is a rapidly advancing field.
Despite significant progress, most studies to date have focused on specific materials and cavity configurations.
Here, through a comprehensive group-theoretical analysis, we establish general symmetry rules for cavity materials engineering with linearly polarized cavity photon modes. 
By analyzing the symmetry of the 
Pauli-Fierz Hamiltonian within the long-wavelength approximation,
we provide a classification of the symmetry-breaking patterns induced by cavity modes for all crystallographic point groups. 
The power of this framework is then demonstrated by quantum-electrodynamical density functional theory calculations. 
In particular, we explain the distinct cavity-induced lifting of band degeneracies in cubic BaTiO$_3$ for different cavity mode configurations, and the cavity-modified infrared and Raman spectra of monolayer MoS$_2$ due to symmetry breaking.
Our results highlight the central role of symmetry
in cavity materials engineering and provide general guidelines for future studies in this field.
\end{abstract}

\maketitle


\section{Introduction}

Over the past decades, controlling material properties through light-matter interaction has emerged as
an active field in condensed matter physics~\cite{ultrafast_rmp_2021, peizhe_ultrafast_NRP,angel_nature_perspective,ultrafast_NRM_2023}. In particular, laser pulses are widely used to drive intriguing non-equilibrium phenomena such as light-induced superconductivity~\cite{light_induced_sc_science_2011,Cavalleri_ultrafast_ybco_2016,light_induced_sc_magnetic_nature}, phase transitions~\cite{cavalleri_vo2_prl_2001,johnson2023ultrafast,Zhiyang_Zeng_photon_chiral}, and Floquet states~\cite{Oka_floquet_review,Lindner2020}. 
Although laser pulses offer an effective route to modifying material properties, they generally access only transient non-equilibrium states with short lifetimes~\cite{ultrafast_rmp_2021,peizhe_ultrafast_NRP} and can also cause heating~\cite{ultrafast_rmp_2021}, which hinders their widespread use for long-lived materials control.

In contrast, cavity materials engineering aims to modify material properties in equilibrium through structured vacuum fluctuations inside a cavity~\cite{angel_nature_perspective,schlawin2022cavity, garcia2021manipulating,lu2025cavity_review,cavity_review_2026}. This approach offers a promising route beyond the limitations of ultrafast control, enabling long-lived modifications of the ground state without external energy pumping. Recent experiments have demonstrated that coupling materials to optical cavities can significantly alter a broad range of properties, including superconductivity~\cite{cavity_SC_jcp,Itai_cavity_SC}, charge density waves~\cite{TaS2_cavity_cdw_nature}, quantum Hall effect~\cite{appugliese2022breakdown,cavity_QH_prx}, and chemical reaction rates~\cite{cavity_reaction_science_2019,cavity_reaction_science_2023}. On the theoretical side, based on quantum electrodynamics (QED), a variety of theoretical approaches have been developed to describe light-matter interactions in optical cavities. Few-state models, such as the quantum Rabi model~\cite{qRabi_review}, Jaynes-Cummings model~\cite{jc_model_book}, Dicke model~\cite{dicke_model_review} and their generalizations \cite{eckhardt2022quantum,passetti_cavity_2023} provide fundamental insights into cavity QED but are limited in describing the light-matter coupling in complicated, realistic systems. At the other end, many-body approaches like QED density matrix renormalization group~\cite{qed_dmrg}, quantum Monte Carlo~\cite{qed_qmc}, configuration interaction~\cite{cavity_CIS,cavity_active_space_CI} and coupled cluster~\cite{qed_cc_prx_2020,cavity_CC_IPEA} methods are much more accurate but suffer from poor scalability, which limits their application. Compared with these approaches, quantum-electrodynamical density functional theory (QEDFT) offers a more favorable balance between accuracy and computational cost, and provides a practical first-principles framework to calculate light-matter interaction in complex molecules and solid state materials~\cite{ruggenthaler2014_prl_qedft, flick_qedft_oep, flick_qedft_fluct_dissip,pf_QEDFT_pnas,I-te_pxLDA,flick2017atoms,aiqed_ruggi_chemical_review,lu2025cavity_review,flick_qedft_qMBD}. Recently, several approximations to the QEDFT exchange-correlation functional have been developed, including the optimized effective potential~\cite{flick_qedft_oep}, photon many-body dispersion~\cite{flick_qedft_qMBD}, and the photon exchange local density approximation (pxLDA)~\cite{I-te_pxLDA, benshu_qedft}. Within the QEDFT framework, a variety of intriguing behaviors such as enhancement of superconductivity~\cite{I-te_mgb2, omid_sc_cavity}, manipulation of topological phases~\cite{Dongbin_HgTe_cavity} and electronic structures~\cite{Hang_qedft_2D, benshu_qedft} with cavity modes have been proposed and investigated.

While experiments and numerical simulations are indispensable for exploring light-matter interaction, a group-theory-based symmetry analysis provides a complementary route to establish a conceptually clear perspective. On the one hand, it enables 
a categorization of possible phenomena prior to heavy numerical computations or complex experiments. On the other hand, it can be used to gain clear physical insight into the mechanisms underlying observed results~\cite{dresselhaus_group_theory,jones_group_theory}.
For example, the energy-level degeneracies are related to the dimensions of the irreducible representations (irreps) of the system's symmetry group~\cite{dresselhaus_group_theory}. Likewise, characteristic peaks in optical and vibrational spectra are largely determined by selection rules. These features are naturally captured by group theory, but are often much more difficult and computationally demanding to extract from first-principles calculations.
In recent years, symmetry analysis has been successfully applied to a variety of ultrafast phenomena~\cite{ofer_light_symmetry_2026,fan2025floquet,li2026first}.
In the field of cavity QED, Jiang {\it et al.}~\cite{QD_Jiang_quantum_atmosphere,QD_Jiang_cavity_chiral} and Cheng {\it et al.}~\cite{xinle_cavity_axion} showed that the symmetry of a material can be transmitted to its vicinity via quantum fluctuations and affect nearby molecules or materials. Jayachandran {\it et al.}~\cite{cavity_molecule_symmetry} interpreted cavity-modified molecular reaction rates under vibrational strong coupling through a phenomenological selection rule. Lin {\it et al.} recently showed that a split-ring resonator can induce spontaneous symmetry breaking in a moir{\'e} superlattice~\cite{lin_spontaneous_2026}. 
Despite these advances, 
most existing studies have focused on model Hamiltonians for specific materials or molecules,
while a comprehensive and systematic group-theoretical framework for cavity materials engineering from first principles is still lacking.

In this work, we analyze the symmetry of the full Pauli-Fierz (PF) Hamiltonian within the long-wavelength approximation, the effective photon-free (pf) QED Hamiltonian, and the pxLDA QEDFT functional. 
This allows us to establish a complete characterization of the symmetry reductions induced by cavity modes polarized along characteristic directions for all 32 crystallographic point groups (PGs). 
We demonstrate the power of this framework by explaining the distinct electronic structure modifications obtained in QEDFT calculations of cubic BaTiO$_3$ for different cavity mode polarization directions.
Furthermore,  we show that 
coupling monolayer MoS$_2$ to a linearly polarized cavity mode gives rise to new infrared (IR) and Raman spectral features.

The paper is organized as follows. In Sec.~\ref{sec: qed hamiltonian}, we introduce the full PF Hamiltonian, the pf QED Hamiltonian, and the pxLDA functional. Section~\ref{sec: symmetry analysis} presents the group-theoretical analysis and the resulting symmetry-broken subgroups. In Secs.~\ref{sec: BTO} and \ref{sec: mos2 vibration}, we demonstrate this framework by QEDFT calculations of the cavity-altered band structure of BaTiO$_3$ and vibrational spectra of MoS$_2$. Finally, in Sec.~\ref{sec: discussion}, we discuss several promising directions stemming from cavity-induced symmetry breaking and in Sec.~\ref{sec: conclusion} we summarize our results.

\section{Methodology}
\label{sec: methodology}

\subsection{The Pauli-Fierz Hamiltonian and photon-free QEDFT}
\label{sec: qed hamiltonian}

In this section, we briefly introduce the PF Hamiltonian, the pf QED Hamiltonian and the pxLDA QEDFT functional.
For this purpose, we start from the PF Hamiltonian of non-relativistic QED in the Coulomb gauge~\cite{pf_QEDFT_pnas,lu2025cavity_review}:
\begin{eqnarray}
\label{eq: PF original}
    \hat{H}_{\rm PF} =&& \frac{1}{2} \sum^{N_e}_{j=1} \Big ( -{\mathrm i}\nabla_j + \frac{1}{c} \hat{\bf A} \Big)^2 
    + \sum^{N_e}_{j=1} v({\bf r}_j) \nonumber \\
    &&+ \frac{1}{2}\sum^{N_e}_{i\neq j} w({\bf r}_i, {\bf r}_j) + \sum^{M_p}_{\alpha = 1} \omega_\alpha \Big ( \hat{a}^\dagger_\alpha \hat{a}_\alpha + \frac{1}{2} \Big ) ,
\end{eqnarray}
where $v({\bf r}_j)$ and $w({\bf r}_i, {\bf r}_j)$ denote the external potential and the electron-electron Coulomb interaction; 
$N_e$ and $M_p$ denote the number of electrons in the unit cell and the number of photon modes;
$\hat{\bf A}$, $\hat{a}$, and $\hat{a}^\dagger$ denote the vector potential, photon annihilation, and photon creation operator, respectively.
In the long-wavelength approximation, the vector potential takes the form
\begin{eqnarray}
    \hat{\bf A} = c \sum^{M_p}_{\alpha = 1} \lambda_\alpha {\bm \epsilon}_\alpha \frac{1}{\sqrt{2\omega_\alpha}} (\hat{a}^\dagger_\alpha + \hat{a}_\alpha) .
    \label{eq: vector potential}
\end{eqnarray}
Here, $\lambda_\alpha$, $\omega_\alpha$, and ${\bm \epsilon}_\alpha$ denote the coupling strength, frequency, and polarization direction of photon mode $\alpha$, respectively.
Note that Eq.~(\ref{eq: PF original}) assumes that the photon frequency is far off resonance with the nuclear vibrations, so that the conventional Born-Oppenheimer (BO) approximation remains valid and the photonic and electronic degrees of freedom can be treated together~\cite{benshu_qedft}. In the opposite regime, the photon modes should be treated on the same footing as the nuclear degrees of freedom, leading to the so-called cavity BO approximation~\cite{flick2017atoms,flick_cavity_BO, Jeremy_cavity_BO}.
The diamagnetic term $\hat{\bf A}^2$ in Eq.~(\ref{eq: PF original}) can then be absorbed into the photon degrees of freedom by a Bogoliubov transformation~\cite{pf_QEDFT_pnas,I-te_pxLDA}, resulting in an equivalent Hamiltonian written in terms of dressed photon modes as
\begin{eqnarray}
    \hat{\tilde{H}}_{\rm PF} &=& -\frac{1}{2} \sum^{N_e}_{j=1} \nabla^2_j +  \sum^{N_e}_{j=1} v({\bf r}_j) + \frac{1}{2}\sum^{N_e}_{i\neq j} w({\bf r}_i, {\bf r}_j) \nonumber \\
    &&+ \frac{1}{c} \hat{\tilde{ {\bf A} }} \cdot \hat{ {\bf \Pi} } + \sum^{M_p}_{\alpha = 1} \tilde{\omega}_\alpha \Big ( \hat{\tilde{a}}^\dagger_\alpha \hat{\tilde{a}}_\alpha + \frac{1}{2} \Big ) .
    \label{eq: HPF dressed}
\end{eqnarray}
Here, the vector potential $\hat{\tilde{\bf A}}$ is defined in terms of dressed photon modes:
\begin{eqnarray}
    \hat{\tilde{ {\bf A}} } = c \sum_\alpha \tilde\lambda_\alpha \tilde{\bm \epsilon}_\alpha \frac{1}{\sqrt{2 \tilde{\omega}_\alpha}} (\hat{\tilde{a}}^\dagger_\alpha + \hat{\tilde{a}}_\alpha) .
    \label{eq: vector potential bogo}
\end{eqnarray}
For orthogonal photon modes, one has $\tilde{\lambda}_\alpha = \lambda_\alpha$, $\tilde{\bm \epsilon}_\alpha = {\bm \epsilon}_\alpha$, and $\tilde{\omega}_{\alpha}^2 = \omega_\alpha^2 + N_e\lambda_\alpha^2$. The operator $\hat{\bf \Pi}$ denotes the total electron momentum, i.e., $\hat{\bf \Pi}= -{\rm i}\sum_j^{N_e} \nabla_j = \sum_j^{N_e} {\bf p}_j$.
Strictly speaking, the complete basis for representing Eq.~(\ref{eq: HPF dressed}) is the tensor product of the matter Hilbert space and the Fock space of dressed photons, i.e. $\mathcal H_e\otimes\mathcal H_{\rm ph}$, where $\mathcal H_{\rm ph}={\rm span}\{\ket{\tilde0},\ket{\tilde1},\ket{\tilde2}, \cdots\}$. 

We now analyze the symmetry of the PF Hamiltonian.  For a spatially homogeneous single cavity mode, Eq.~(\ref{eq: HPF dressed}) can be  rearranged as
\begin{eqnarray}
    {H} = H_e + \tilde{\omega}\hat{\tilde{a}}^\dagger\hat{\tilde{a}} + \tilde{X}_{\tilde{\bm \epsilon}}(\hat{\tilde{a}}^\dagger+\hat{\tilde{a}})
    \label{eq: dressed PF}
\end{eqnarray}
where 
\begin{eqnarray}
    \tilde{X}_{\tilde{\bm \epsilon}} = \frac{\tilde{\lambda}}{\sqrt{2\tilde{\omega}}}\tilde{{\bm \epsilon}}\cdot \hat{\bf \Pi}
\end{eqnarray}
and a constant energy shift has been omitted.
Let $U_g$ denote a symmetry operation of the bare electronic Hamiltonian $H_e$, represented in real space by the orthogonal matrix $R_g$. Since $U_g{\bf p}U_g^{-1} = R_g {\bf p}$,
the electronic light-matter coupling transforms as
\begin{eqnarray}
    U_g\tilde{X}_{\tilde{\bm \epsilon}}U_g^{-1} &=& \frac{\tilde{\lambda}}{\sqrt{2\tilde{\omega}}}\tilde{{\bm \epsilon}}\cdot R_g\hat{\bf \Pi}  \nonumber \\
    &=&\frac{\tilde{\lambda}}{\sqrt{2\tilde{\omega}}}R^T_g\tilde{{\bm \epsilon}}\cdot \hat{\bf \Pi} = \tilde{X}_{R^T_g\tilde{\bm \epsilon}}.
\end{eqnarray}
As a result, for a single cavity mode, if $R^T_g\tilde{\bm \epsilon}$ is not parallel to $\tilde{\bm \epsilon}$, the symmetry operation transforms the light-matter interaction into one associated with a different cavity polarization. In this case, $U_g$ is broken by the cavity coupling.

If instead $R^T_g\tilde{\bm \epsilon} \parallel \tilde{\bm \epsilon}$, there are two possibilities. For $R^T_g\tilde{\bm \epsilon} = \tilde{\bm \epsilon}$, the interaction term is invariant and therefore $(U_g\otimes\mathbb{I}) H (U_g\otimes\mathbb{I})^{-1}=H$. For $R^T_g\tilde{\bm \epsilon} = -\tilde{\bm \epsilon}$, the transformation requires additional consideration. 
In this case, the electronic coupling changes sign, $U_g \tilde{X}_{\tilde{\bm \epsilon}} U_g^{-1} = - \tilde{X}_{\tilde{\bm \epsilon}}$. The full PF Hamiltonian is therefore not invariant under $U_g\otimes\mathbb{I}$. However, introducing the photon-parity operator $B_{\rm ph}=\exp(i\pi\hat{\tilde{a}}^\dagger\hat{\tilde{a}})$ which satisfies $B_{\rm ph}(\hat{\tilde{a}}^\dagger + \hat{\tilde{a}})B^{-1}_{\rm ph}=-(\hat{\tilde{a}}^\dagger + \hat{\tilde{a}})$, we find that the combined operation $\mathcal{ D}_g = U_g \otimes B_{\rm ph}$ leaves the full PF Hamiltonian invariant, $\mathcal{ D}_gH\mathcal{ D}^{-1}_g=H$. Importantly, although the symmetry of the full electron-photon system is then the joint symmetry
$\mathcal{D}_g$, the corresponding symmetry $U_g$ is retained in the reduced electronic subsystem. To demonstrate this, consider the reduced electronic density matrix $\rho_e = {\rm Tr}_{\rm ph} \rho$, where ${\rm Tr}_{\rm ph}$ denotes the partial trace over the photon degrees of freedom and $\rho$ is the equilibrium density matrix of the coupled electron-photon system. For a nondegenerate equilibrium ground state of the coupled system in a dark cavity or more generally for a state that preserves the joint symmetry, one has $\mathcal{D}_g \rho \mathcal{D}_g^{-1} = \rho$. It then follows that
\begin{eqnarray}
    U_g \rho_e U_g^{-1} &=&  {\rm Tr}_{\rm ph}\left[ U_g \rho U_g^{-1}\right] \nonumber \\
    &=& {\rm Tr}_{\rm ph}\left[ U_g \mathcal{D}_g^{-1}\rho\mathcal{D}_g U_g^{-1}\right] \nonumber \\
    &=& {\rm Tr}_{\rm ph}\left[ {B^{-1}_{\rm ph}}\rho B_{\rm ph} \right] = {\rm Tr}_{\rm ph} \rho = \rho_e.
\end{eqnarray}
Therefore, for a spatially homogeneous single cavity mode, a crystal symmetry operation $U_g$ remains a symmetry of the electronic subsystem whenever $R^T_g\tilde{\bm \epsilon} = \pm \tilde{\bm \epsilon}$. 
Conversely, operations that rotate the cavity polarization axis into a different direction are broken. We emphasize that this symmetry analysis assumes a symmetry-preserving state. A phase-selected coherent photon state, such as that associated with a laser, generally does not preserve $\mathcal{D}_{g}$ symmetry and may therefore exhibit different symmetry-breaking behavior.

In the high-frequency limit, i.e., when the dressed photon frequency is much larger than the relevant electronic transition frequencies, different photon sectors become off-resonant. The PF Hamiltonian can then be downfolded onto the zero photon sector, effectively eliminating the explicit photonic degrees of freedom by a Schrieffer-Wolff transformation~\cite{Simone_sto_pnas,Hang_qed_hf}:\begin{eqnarray}
\label{eq: full SW}
        H_{\rm eff} &=& H_e - \frac{\tilde{X}_{\tilde{\bm \epsilon}}^2}{\tilde{\omega}} + \frac{1}{2\tilde{\omega}^2}\left[ \tilde{X}_{\tilde{\bm \epsilon}}, \left[ H_e, \tilde{X}_{\tilde{\bm \epsilon}}\right] \right] \nonumber \\
    &-& \frac{1}{2\tilde{\omega}^3}\left\{\tilde{X}_{\tilde{\bm \epsilon}}, \left[ H_e,\left[ H_e, \tilde{X}_{\tilde{\bm \epsilon}}\right] \right] \right\}+ O({\tilde{\omega}^{-4}}).
\end{eqnarray}
Nonvanishing corrections to the electronic Hamiltonian arise from even-order virtual processes in which a virtual photon is created and subsequently annihilated. Clearly, Eq.~(\ref{eq: full SW}) inherits the symmetry constraints of the full PF Hamiltonian. In the full PF Hamiltonian, the sign change of $\tilde{X}_{\tilde{\bm \epsilon}}$ is compensated by photon parity. After eliminating the photons, the same symmetry is manifest in the quadratic interaction $\tilde{X}_{\tilde{\bm \epsilon}}^2$ in the downfolded electronic Hamiltonian.
If we take the leading order correction in Eq.~(\ref{eq: full SW}), the PF Hamiltonian of a time-independent system reduces to
\begin{eqnarray}
\label{eq: photon-free H}
    \hat{H}^{\rm pf} &=& -\frac{1}{2} \sum^{N_e}_{i=1} \nabla^2_i +  \sum^{N_e}_{i=1} v({\bf r}_i) + \frac{1}{2}\sum^{N_e}_{i\neq j} w({\bf r}_i, {\bf r}_j) \nonumber \\
    && + \sum^{M_p}_{\alpha = 1} \frac{\tilde{\omega}_\alpha}{2} - \sum^{M_p}_{\alpha = 1} \frac{\tilde{\lambda}_\alpha^2}{2\tilde{\omega}_\alpha^2} \Big ( \hat{\bf \Pi} \cdot \tilde{{\bm \epsilon}}_\alpha \Big )^2 .
\end{eqnarray}
This reduced Hamiltonian is often referred to as the effective pf QED Hamiltonian~\cite{pf_QEDFT_pnas}.  
From the above expression, the effective light-matter interaction term for a cavity mode $\alpha$ is proportional to
\begin{eqnarray}
\label{eq: symmetry of photon-free qed}
     \hat{H}^{\rm pf}_I\sim ({\bf \hat{\bf \Pi}} \cdot \tilde{\bm \epsilon}_\alpha)^2 .
\end{eqnarray}

Inspired by the Kohn-Sham framework of density functional theory (DFT), one can likewise construct an auxiliary non-interacting system that reproduces the exact electron density of the pf QED Hamiltonian Eq.~(\ref{eq: photon-free H}), which incorporates the vacuum light-matter interaction. 
The corresponding electron-photon exchange-correlation potential can be constructed, for example, from the local-force equation~\cite{force_balance_eq_2019,force_balance_eq_2024}. Following this approach, Lu {\it et al.}~\cite{I-te_pxLDA} recently developed a local-density approximation to the electron-photon exchange potential, denoted as $v_{\rm pxLDA}$. 
In this framework, the pxLDA potential is obtained by solving the Poisson equation
\begin{eqnarray}
\label{eq: poisson pxlda}
    \nabla^2 v_{\rm pxLDA}({\bf r}) = - \sum_\alpha \frac{2\pi^2 \tilde{\lambda}^2_\alpha}{\tilde{\omega}^2_\alpha}
    \left \{ (\tilde{\bm \epsilon}_\alpha \cdot \nabla)^2 \left [  \frac{\rho({\bf r})}{2V_d}\right]^{\frac{2}{d}}\right\}, \nonumber \\
\end{eqnarray}
where $d$ is the dimensionality of the system, $V_d$ is the corresponding unit cell volume, and $\rho({\bf r})$ is the electron density.
By transforming Eq.~(\ref{eq: poisson pxlda}) into $\bf k$-space, we obtain
\begin{eqnarray}
\label{eq: v_pxlda in k}
    v_{\rm pxLDA} ({\bf k}) = - \sum_\alpha \frac{2\pi^2 \tilde{\lambda}^2_\alpha}{\tilde{\omega}^2_\alpha (2V_d)^{2/d}} \frac{( {\bf k} \cdot \tilde{\bm \epsilon}_\alpha )^2}{{\bf k}^2} g[\rho]({\bf k}) ,
\end{eqnarray}
where $g[\rho]({\bf k})$ is the Fourier transform of $\rho^{2/d}({\bf r})$.
Since {${\bf k}^2$} and $g[\rho]({\bf k})$ are invariant under the point-group symmetry operations of the material, the symmetry reduction induced by a single cavity photon mode $\alpha$ is governed by
\begin{eqnarray}
\label{eq: pxlda_coupling}
    v_{\rm pxLDA} ({\bf k}) \sim ({\bf k} \cdot \tilde{\bm \epsilon}_\alpha )^2  .
\end{eqnarray}
Note that in a fully self-consistent calculation including cavity modes, $g[\rho]({\bf k})$ adapts and inherits the reduced symmetry of the coupled system. However, that reduced symmetry is determined by the kernel $({\bf k} \cdot \tilde{\bm \epsilon}_\alpha )^2$. Although $g[\rho]({\bf k})$ can reflect the symmetry breaking introduced by the cavity, it cannot further lower the symmetry beyond what is imposed by the kernel.
This result is fully consistent with the symmetry of the pf QED Hamiltonian Eq.~(\ref{eq: symmetry of photon-free qed}) as well as the full PF Hamiltonian discussed above.
Such agreement is expected, since the pxLDA functional provides a density-functional approximation to the underlying pf QED Hamiltonian. A similar quadratic dependence on the cavity modes also appears in QED Hartree-Fock theory~\cite{Hang_qed_hf}, which is also based on the pf QED Hamiltonian.
From Eqs.~(\ref{eq: photon-free H}) and (\ref{eq: poisson pxlda}), the strength of the cavity-induced effect in the pf QED framework is determined by the prefactor $\tilde{\lambda}^2/\tilde{\omega}^2$. For a Fabry-P{\'e}rot cavity, the bare coupling strength $\lambda$ is determined by the mode volume $V_{\rm mode}$ via $\lambda=\sqrt{4\pi N_{\rm cell}/V_{\rm mode}}$ ~\cite{benshu_qedft}. 
For a single cavity mode, one obtains
\begin{eqnarray}
    \frac{\tilde{\lambda}^2}{\tilde{\omega}^2} = 
\frac{4\pi N_{\rm cell}}{\omega^2V_{\rm mode}+4\pi N_{\rm cell}N_e}.
\end{eqnarray}
Therefore, achieving a larger $\tilde{\lambda}^2/\tilde{\omega}^2$ generally requires a small $V_{\rm mode}$, i.e., a strongly confined cavity field, a lower cavity photon frequency $\omega$, and a smaller number of electrons per unit cell $N_e$.

The effect of a cavity on electronic states within the pf QED framework admits a natural geometric interpretation.
As shown in Eq.~(\ref{eq: photon-free H}), the cavity light-matter interaction contributes an effective quadratic correction to the kinetic term. This can be equivalently interpreted as a modification of the canonical momentum metric $\underline{h}$ if we collect all terms related to the electron momentum in the Hamiltonian as
\begin{eqnarray}
         -\frac{1}{2} \sum^{N_e}_{i=1} \nabla^2_i -\sum_\alpha \frac{\tilde{\lambda}_\alpha^2}{2\tilde{\omega}_\alpha^2} (\hat{\bf\Pi} \cdot \tilde{\bm \epsilon}_\alpha)^2 = \frac{1}{2}{\bf P}^T \underline{h}^T\underline{h} {\bf P} ,
\end{eqnarray}
where ${\bf P}=({\bf p}_1, {\bf p}_2, \dots, {\bf p}_{N_e})^T$ is the $3N_e$-dimensional vector consisting of all electronic momenta, and the superscript $T$ denotes transposition. 
This form shows that photonic fluctuations in a linearly polarized dark cavity effectively renormalize the metric experienced by the electrons.
Equivalently, this can also be viewed as renormalizing the electron mass along the photon polarization direction $\tilde{\bm \epsilon}$, as discussed in the literature~\cite{free_electron_gas_cqed,pf_QEDFT_pnas,aiqed_ruggi_chemical_review,qed_mass_ruggi_prr_2025}.

\subsection{Symmetry-broken Subgroups}
\label{sec: symmetry analysis}

\begin{table*}[ht!]
    \centering
    \begin{tabular}{l l}
        \toprule
        Preserved Symmetries & Broken Symmetries \\
         \midrule
         \cmark \ Inversion symmetry $\mathcal{P}$ & \xmark \ 3-, 4- and 6-fold rotation axes $\mathcal C_n$ not parallel to $\tilde{\bm \epsilon}$\\
         \cmark \ Time-reversal symmetry $\mathcal{T}$ & \xmark \ Combined symmetry $\mathcal C_n \mathcal T$ when $\mathcal C_n$ is broken\\
         \cmark \ Combined symmetry $\mathcal{PT}$ & \xmark \ Combined symmetry $\mathcal C_n{\mathcal P}$ when $\mathcal C_n$ is broken \\
         \cmark \ 2-fold rotation axes $\mathcal{C}_2$ perpendicular to $\tilde{\bm \epsilon}$ & \xmark \ Mirror planes neither perpendicular nor containing $\tilde{\bm \epsilon}$\\
         \cmark \ All rotation axes parallel to $\tilde{\bm \epsilon}$ & 
         \\
         \cmark \ Mirror planes perpendicular to or containing $\tilde{\bm \epsilon}$ & 
        \\
        \cmark \ Translational symmetry $\bm \tau$
        \\
         \bottomrule
    \end{tabular}
    \caption{Summary of representative symmetry operations preserved and broken under coupling to an $\tilde{\bm \epsilon}$-polarized cavity mode,  or equivalently to two orthogonal cavity modes polarized perpendicular to $\tilde{\bm \epsilon}$. }
    \label{tab:broken_symmetries}
\end{table*}

In the last section, we demonstrated that the electronic subsystem of the nondegenerate ground state of the full PF Hamiltonian, the pf QED Hamiltonian, and the pxLDA functional obey the same symmetry-breaking rules. 
In the remainder of this paper, we will carry out further symmetry analysis using Eq.~(\ref{eq: symmetry of photon-free qed}) as a representative, without loss of generality.
In comparison with Eq.~(\ref{eq: symmetry of photon-free qed}), the light-matter coupling to a given classical electromagnetic field in the velocity gauge takes the form:
\begin{eqnarray}
    \hat H_I^{\rm cl} = \frac{q}{mc} \hat {\bf \Pi} \cdot {\bf A}(t) ,
\end{eqnarray}
where $q$ and $m$ are the charge and mass of the electron, and ${\bf A}(t)$ is the externally prescribed vector potential. 
For a field polarized along $\bm{\epsilon}$, the corresponding linear coupling is governed by $\hat{\bf \Pi}\cdot\bm{\epsilon}$. By contrast, the leading effective interaction induced by vacuum fluctuations in the pf QED Hamiltonian is quadratic in the electronic momentum:
\begin{eqnarray}
    \hat H_I^{\rm pf} &\sim& (\hat{\bf\Pi} \cdot \tilde{\bm \epsilon})^2, \\
    \hat H_I^{\rm cl} &\sim& ({\hat {\bf \Pi}} \cdot \bm \epsilon).
\end{eqnarray}
This distinction arises from the different treatments of the electromagnetic field. In an equilibrium dark cavity, the photon field is a dynamical quantum degree of freedom, and the light-matter coupling preserves the joint electron-photon symmetries, as demonstrated in Sec.~II A. Upon eliminating the photonic degrees of freedom in the high-frequency limit, the leading correction to the electronic Hamiltonian arises from second-order virtual-photon processes and is therefore quadratic in the light-matter coupling operator. In contrast, a prescribed classical field with a fixed phase acts as an external perturbation and generally introduces a linear momentum-dependent interaction.

From a symmetry perspective, one of the key distinctions between these two light-matter coupling Hamiltonians is that the interaction in a dark cavity cannot break inversion symmetry ($\mathcal P$), whereas an external drive can. For the pf QED interaction term one has:
\begin{eqnarray}
    {\mathcal P} \hat{H}_I^{\rm pf} \mathcal P^\dagger= \sum_\alpha \frac{\tilde{\lambda}^2_\alpha}{2\tilde{\omega}_\alpha^2}(-\hat{\bf \Pi}\cdot \tilde{\bm \epsilon}_\alpha)^2 = \hat{H}_I^{\rm pf} .
\end{eqnarray}
By contrast, a classical electromagnetic field introduces a linear coupling term, which can break {$\mathcal{P}$}. 
Similarly, linearly polarized cavity modes do not break time-reversal symmetry ($\mathcal{T}$), and therefore preserve $\mathcal{PT}$ symmetry as well.
More generally, $\hat{H}^{\rm pf}_I$ preserves all symmetry operations that leave 
$(\hat{\bf \Pi}\cdot\tilde{\bm \epsilon}_\alpha)^2$
invariant, and breaks those that do not.
Based on this principle, we summarize the crystal symmetry operations that are preserved or broken by a linearly polarized cavity mode in Tab.~\ref{tab:broken_symmetries}. 
For instance, a cavity mode polarized along the $x$ direction, i.e., $\tilde{\bm \epsilon}=(1,0,0)$, induces a coupling term
\begin{eqnarray}
    \hat{H}_I^{\rm pf} \sim \Pi_x^2.
\end{eqnarray}
Consequently, any rotation that changes the $\Pi_x$ component, such as a $\mathcal C_3$, $\mathcal C_4$, or $\mathcal C_6$ rotation about an axis perpendicular to the $x$ direction, is broken. By contrast, symmetry operations such as a $\mathcal C_2$ rotation about an axis perpendicular to $x$ or the mirror operation $\mathcal{M}_x$, which map $\Pi_x \to -\Pi_x$, preserve the quadratic form and thus remain valid symmetries. 
Moreover, preserving $\Pi_x^2$ is equivalent to preserving $\Pi_y^2 + \Pi_z^2$. Therefore, the symmetry breaking pattern induced by a single cavity mode polarized along $\tilde{\bm \epsilon}$ is the same as that {induced} by two orthogonal cavity modes with the same mode frequency and coupling strength polarized in the plane perpendicular to $\tilde{\bm \epsilon}$. 

\begin{table}
\centering
\begin{tabular}{ c c@{\hspace{0.7cm}} c c}
\toprule
 \multirow{2}{*}{PG} & {Cavity Mode} & \multirow{2}{*}{PG} & {Cavity Mode}\\
 &  along $x$-/$y$-axis & &  along $x$-/$y$-axis \\
 \midrule
 C$_{1}$ & $\circ$ & D$_{4h}$ & D$_{2h}$\\
 C$_i$ & $\circ$ & C$_{3}$ & C$_1$\\
C$_2$ & $\circ$ & S$_6$ & C$_i$\\
C$_s$ & $\circ$ & D$_3$ & C$_2$\\
C$_{2h}$ & $\circ$& C$_{3v}$ & C$_s$\\
D$_2$ & $\circ$ & D$_{3d}$ & C$_{2h}$\\
C$_{2v}$ & $\circ$ & C$_6$ & C$_2$\\
D$_{2h}$ & $\circ$ & C$_{3h}$ & C$_s$\\
D$_{2d}$ & C$_{2v}$ & C$_{6h}$ & C$_{2h}$\\

C$_4$ & C$_2$ & D$_6$ & D$_{2}$ \\
S$_{4}$ & C$_{2}$ & C$_{6v}$ & C$_{2v}$\\
C$_{4h}$ & C$_{2h}$ & D$_{3h}$ & C$_{2v}$\\
D$_{4}$ & D$_{2}$ & D$_{6h}$ & D$_{2h}$ \\
C$_{4v}$ & C$_{2v}$ & \\
\bottomrule

\end{tabular}
\caption{Symmetry-broken subgroups of uniaxial PGs induced by a single linearly polarized cavity mode, assuming that the principal symmetry axis of the cavity-free system lies along $z$. The corresponding coordinate systems are shown in Fig.~\ref{fig:PG coordinates}. The symbol $\circ$ denotes that the PG remains unchanged under cavity coupling. Two orthogonal cavity modes in the ($x+z$) or ($y+z$) configuration generate the same symmetry-broken subgroup. }
\label{tab: C and D group}
\end{table}

\begin{figure}[ht!]
    \centering
    \includegraphics[width=1\linewidth]{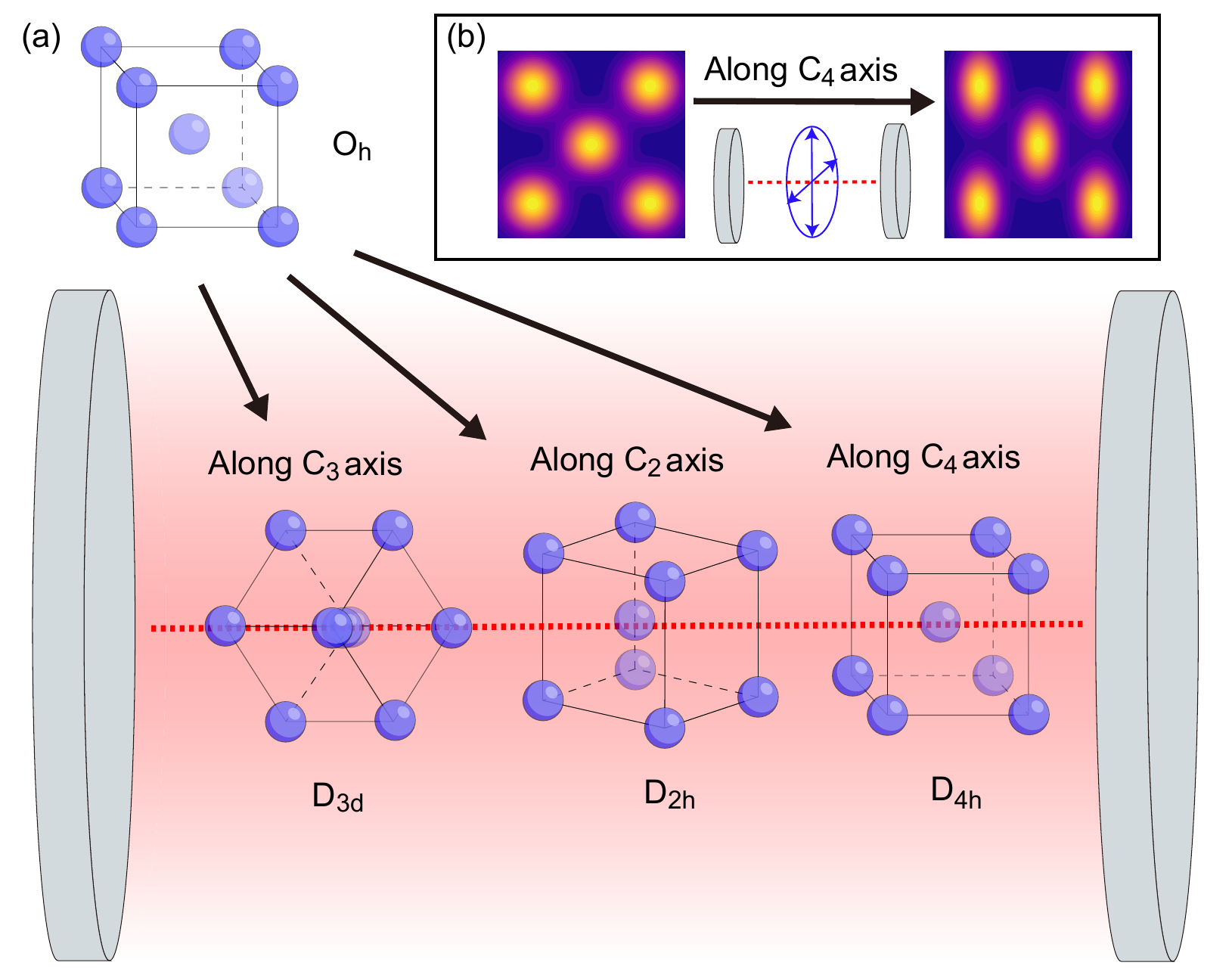}
    \caption{
    (a) Schematic illustration of a body-centered cubic cell placed inside a Fabry-P\'{e}rot cavity. Different cavity-mode polarizations relative to the crystal can be realized by rotating the crystal within the cavity, leading to different symmetry-reduced subgroups. (b) Schematic of the electron density redistribution in a body-centered cubic cell coupled to two cavity modes polarized perpendicular to its $\mathcal C_4$ axis.}
    \label{fig:Animate_cavity}
\end{figure}

Based on the symmetry analysis above, we now discuss the symmetry-breaking behavior of all 32 crystallographic PGs under coupling to cavity modes. 
Specifically, we determine the corresponding subgroup for the original PG by removing the symmetry operations listed in Tab.~\ref{tab:broken_symmetries} that are broken by cavity modes.
Before presenting the results, we briefly summarize the effective cavity mode configurations for several cavity setups.
In principle, the continuum of photon modes should be considered in cavity QED calculations~\cite{gruner_mqed_1996,lindel_mqed_2021,mark_mqed_dft_nc,Pantazopoulos_pf_qed,Eugene_continuum_mode_cQED}. However, their effect can be approximated by several effective photon modes in the high-frequency or strong-coupling limit~\cite{lu2025cavity_review,Mark_effective_single_mode_2025}.
In a standard Fabry-P\'{e}rot optical cavity, as illustrated in Fig.~\ref{fig:Animate_cavity}, the photon-mode polarization directions lie in the plane parallel to the mirrors and can be effectively represented by two orthogonal linearly polarized modes~\cite{lu2025cavity_review}. In other common cavity setups, such as dielectric planar waveguides~\cite{Dongbin_HgTe_cavity}, split-ring resonators~\cite{schlawin2022cavity,lin_spontaneous_2026}, or subwavelength surface cavities~\cite{Emil_surface_cavity_prr,Chongxiao_niI2}, the cavity field  can be effectively described by a single linearly polarized mode. 
Throughout this paper, we therefore restrict our analysis to either one or two effective photon modes.

\begin{table}[ht!]
\centering
\begin{tabular}{ l *{5}c}
\toprule
\multirow{2}{*}{PG} & \multicolumn{5}{c}{Cavity Mode along axes} \\
 & $\mathcal{C}_2$ & $\mathcal{C}_3$ & $\mathcal{C}_4$ & $\mathcal{C}_2\mathcal{P}$ & $\mathcal{C}_4\mathcal{P}$ \\
 \midrule
         T & D$_2$ & C$_3$ & $\times$ & $\times$ & $\times$ \\
         T$_h$ & D$_{2h}$ & S$_6$ & $\times$ & D$_{2h}$ & $\times$ \\
         O & D$_2$ & D$_3$ & D$_4$ & $\times$ & $\times$\\
         T$_d$ & D$_{2d}$ & C$_{3v}$ & $\times$ & C$_{2v}$ & D$_{2d}$\\
         O$_h$ & D$_{2h}$ & D$_{3d}$ & D$_{4h}$ & D$_{2h}$ & D$_{4h}$\\
         \bottomrule
    \end{tabular}
    \caption{Symmetry-broken subgroups for polyhedral PGs induced by a linearly polarized cavity mode along different high-symmetry axes. The symbol $\times$ indicates that the corresponding symmetry operation is absent in this group. The high-symmetry axes are illustrated in Fig.~\ref{fig:PG coordinates}.}
    \label{tab: T and O group}
\end{table}

Tab.~\ref{tab: C and D group} summarizes the symmetry-broken subgroups of all uniaxial PGs under a single linearly polarized cavity mode, assuming that the principal axis is aligned along $z$. The corresponding coordinate systems for PGs with principal axes $\mathcal{C}_4$, $\mathcal{C}_2$, $\mathcal{C}_6$, and $\mathcal{C}_3$ are illustrated in Fig.~\ref{fig:PG coordinates}(a)-(d) of Appendix~\ref{app: PG coordinate}, respectively. 
For uniaxial PGs, the analysis is relatively simple, since there is only one high-symmetry rotation axis.
It is clear from Tab.~\ref{tab:broken_symmetries} that a cavity mode polarized along $z$, or two orthogonal cavity modes with the same $\lambda$ and $\omega$ polarized in the $xy$-plane, for example polarized along $x$ and $y$ (denoted as $(x+y)$), does not break any symmetry. Likewise, Tab.~\ref{tab:broken_symmetries} and the discussion above indicate that, for uniaxial groups, one cavity mode along the $x$ or $y$ direction, and two orthogonal cavity modes with the same $\tilde{\lambda}/\tilde{\omega}$ in the $(x+z)$ or $(y+z)$ configuration generate the same symmetry-broken subgroup. Tab.~\ref{tab: C and D group} is therefore sufficient to describe both one- and two-mode cavity configurations. 
Note that some subgroup assignments depend on the coordinate convention. For instance, for the D$_{2d}$ group, the $x$ and $y$ axes lie within the mirror planes in our convention. Thus, a cavity mode polarized along the $x$ or $y$ axis preserves the mirror planes but breaks the in-plane $C_2$ operations. Therefore the resulting subgroup is C$_{2v}$ in Tab.~\ref{tab: C and D group}. However, with a coordinate convention that these cavity modes instead preserve the in-plane $C_2$ operations and break the mirror planes, the resulting subgroup should be D$_2$ instead.

For polyhedral PGs, the symmetry breaking patterns are more intricate, since multiple inequivalent high-symmetry axes coexist. In this case, it is useful to classify the resulting subgroups by cavity mode polarization directions. 
Figures~\ref{fig:PG coordinates}(e) and (f) illustrate characteristic high-symmetry axes in the polyhedral PGs.
Accordingly, Tab.~\ref{tab: T and O group} summarizes the symmetry-broken subgroups for a single linearly polarized cavity mode along representative high-symmetry axes, while two orthogonal modes with equal $\lambda$ and $\omega$ polarized in the plane perpendicular to a given axis yield the same subgroup. 
As shown there, cavity modes polarized along different axes or crystal planes can realize distinct symmetry reductions, offering substantial flexibility for cavity materials engineering. 
For example, Fig.~\ref{fig:Animate_cavity}(a) shows a body-centered cubic cell with point group ${\rm O}_h$. It possesses $\mathcal C_2$, $\mathcal C_3$, and $\mathcal C_4$ rotational symmetries, along the [110], [111], and [001] directions, respectively. When placed in a Fabry-P\'{e}rot cavity such that these axes are normal to the cavity-mode polarization plane, the corresponding PG is reduced to ${\rm D}_{2h}$, ${\rm D}_{3d}$, and ${\rm D}_{4h}$, respectively. These different symmetry reductions originate from the distinct electron density distortions induced by the pf QED Hamiltonian, as illustrated in Fig.~\ref{fig:Animate_cavity}(b). 

We note that a cavity mode polarized along a generic direction may break all PG symmetries. However, such situations are usually not of primary interest, as materials engineering commonly aims to break selected symmetries while preserving others. In principle, the full symmetry of a nonmagnetic crystal is described by its space group, including translational symmetry and nonsymmorphic operations such as screw axes and glide planes~\cite{dresselhaus_group_theory}. However, the pf QED Hamiltonian in the long-wavelength limit preserves translational symmetry. And our point-group analysis determines the rotational parts of the space-group operations. Consequently, in nonsymmorphic space groups, a screw axis or glide plane survives whenever its associated rotation or mirror operation is preserved.
Further discussion of cavity materials engineering beyond the long-wavelength limit is provided in Sec.~\ref{sec: discussion}.
Note that, in Tabs.~\ref{tab: C and D group} and~\ref{tab: T and O group}, the underlying lattice structure is not specified explicitly, so we use Cartesian directions and rotation axes to indicate the cavity mode directions.
In the following examples of real materials, both Cartesian directions and the corresponding Miller indices are provided to avoid ambiguity.

\section{Applications}

In the previous section, we presented the group-theoretical analysis of the PF Hamiltonian and the effective pf QED Hamiltonian. We now discuss several applications that illustrate the power of symmetry analysis in cavity materials engineering. Computational details are provided in Appendix~\ref{app: computational details}.

\subsection{Band Structures of Cubic BaTiO$_3$}
\label{sec: BTO}

\begin{center}
    \begin{figure*}[ht!]
        \centering
    \includegraphics[width=0.9\linewidth]{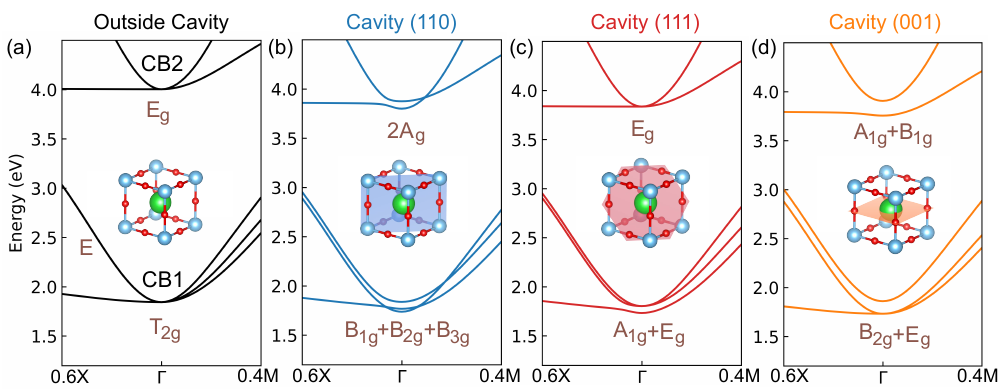}
        \caption{
        (a)-(d) Band structures of cubic BaTiO$_3$ from QEDFT for different cavity photon modes. The $\bf k$-path used in our calculation is $\rm X \ (0.5, 0.0, 0.0)$-$\Gamma \ (0.0, 0.0, 0.0)$-$\rm M \ (0.5, 0.5, 0.0)$, and a zoomed-in near $\Gamma$ and the conduction-band minimum is shown here. The inset illustrates the crystal structure of BaTiO$_3$ and the plane spanned by the two orthogonal cavity modes. The irreps of wavefunctions at $\Gamma$ are also indicated.
        }
        \label{fig:BTO_band}
    \end{figure*}
\end{center}

Cubic BaTiO$_3$ is one of the most widely studied materials in the oxide perovskite family~\cite{BTO_review, oxide_perovskite_review,oxide_perovskite_review_2}, whose point group is O$_h$. Here, we choose it as a concrete example to illustrate cavity-induced symmetry breaking.
Figure~\ref{fig:BTO_band}(a) shows the band structures of pristine cubic BaTiO$_3$ around the conduction band minimum outside the cavity. At the $\Gamma$ point, the first and second conduction-band manifolds, denoted as CB1 and CB2, are threefold and twofold degenerate, respectively. As shown in Fig.~\ref{fig:BTO_band}(a), the wavefunctions of CB1 and CB2 correspond to the $T_{2g}$ and $E_g$ irreps of the O$_h$ group, which are mainly composed of Ti-$3d$ orbitals~\cite{BTO_band_component}.

As summarized in Tab.~\ref{tab: T and O group}, cavity modes polarized along different directions can reduce O$_h$ to different subgroups. 
For cubic BaTiO$_3$, the $(110)$, $(111)$, and $(001)$ crystal planes have normals along the $\mathcal{C}_2$, $\mathcal{C}_3$, and $\mathcal{C}_4$ axes, respectively. Accordingly, two cavity modes polarized in these planes yield the symmetry-broken subgroups D$_{2h}$, D$_{3d}$, and D$_{4h}$, respectively. 
Under these symmetry reductions, the degeneracies of CB1 and CB2 could change. 
First, for cavity modes in the $(110)$ plane, the relevant irreps decompose under the D$_{2h}$ group as
\begin{equation}
\begin{aligned}
\label{eq: Oh to D2h}
    T_{2g} &\to B_{1g} + B_{2g} + B_{3g}, 
    \\
    E_g &\to A_g + A_g.
\end{aligned}
\end{equation}
Since all resulting irreps are one-dimensional, all degeneracies are removed, in agreement with the QEDFT results shown in Fig.~\ref{fig:BTO_band}(b).
Second, for cavity modes in the $(111)$ plane, the irreps decompose under the D$_{3d}$ group as
\begin{equation}
\begin{aligned}
        T_{2g} &\to A_{1g} + E_{g}, 
        \\
    E_g &\to E_g.
\end{aligned}
\end{equation}
In this case, CB1 splits into one nondegenerate band and one doubly degenerate band, but CB2 remains doubly degenerate, as shown in Fig.~\ref{fig:BTO_band}(c). Moreover, for cavity modes in the $(001)$ plane, the irreps decompose under the D$_{4h}$ group as
\begin{equation}
    \begin{aligned}
        T_{2g} &\to B_{2g} + E_{g}, 
        \\
    E_g &\to A_{1g}+B_{1g}.
    \end{aligned}
\end{equation}
Accordingly, as shown in Fig.~\ref{fig:BTO_band}(d), the CB1 again splits into one nondegenerate band and one doubly degenerate band, while the degeneracy of the CB2 is lifted. Besides the $\Gamma$ point, the little-group symmetry at general $\bf k$ points~\cite{dresselhaus_group_theory} is also changed by cavity light-matter interaction. For example, in cubic BaTiO$_3$, the wavefunctions at $\bf k$ points along the high-symmetry path $\rm X - \Gamma$ have C$_{4v}$ little-group symmetry. As shown in Fig.~\ref{fig:BTO_band}(a), the second band along $\rm X - \Gamma$ is doubly degenerate and belongs to the irrep $E$. Owing to the symmetry breaking, this degeneracy is also lifted, as shown in Figs.~\ref{fig:BTO_band}(b)-(d). 
Note that, to show the band splitting more clearly, we use a relatively large coupling ratio $\lambda/\omega=0.2$ in the QEDFT calculations. In experiments, the maximum of this ratio can be on the order of $0.1$~\cite{benshu_qedft,I-te_mgb2}. Degeneracy lifting can also be observed with smaller coupling ratio, since the symmetry-breaking mechanism remains valid.
These results demonstrate that different cavity configurations can lead to distinct symmetry breaking and result in different electronic structure modifications. 

In Appendix~\ref{app: unequal coupling}, we discuss cavity configurations with unequal coupling strengths along different cavity-mode directions.
Furthermore, in Appendix~\ref{app: Dirac cone shift}, we also discuss the momentum shift of the Dirac cone in graphene as another example of cavity-induced electronic structure modification.

\subsection{Vibrational Spectra of Monolayer MoS$_2$}
\label{sec: mos2 vibration}

\begin{center}
    \begin{figure*}[ht!]
        \centering
        \includegraphics[width=1\linewidth]{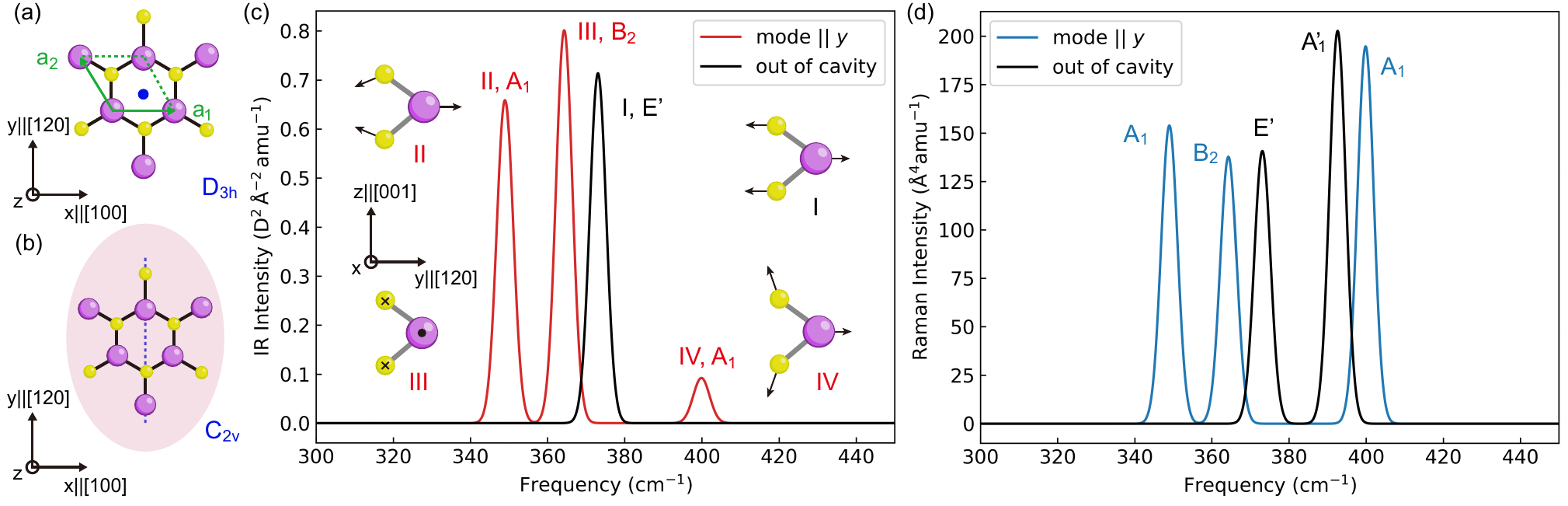}
        \caption{
        (a) Illustration of the crystal structure of MoS$_2$ with the ${\rm D}_{3h}$ point group outside the cavity. The green arrows indicate the lattice vectors used in our calculations, and the blue dot marks the principal axis along the $z$ direction.
        (b) Illustration of MoS$_2$ coupled to a $y$-polarized (along the [120] direction) cavity mode, under which the symmetry is reduced to ${\rm C}_{2v}$. The blue dashed line indicates the new principal axis along the $y$ direction.
        (c) Infrared spectra of MoS$_2$ with and without a cavity mode, calculated using QEDFT. The insets show the vibrational eigenmodes corresponding to the four modes shown in this figure. (d) Same as (c), but for the Raman spectra.}
        \label{fig:MoS2_IR}
    \end{figure*}
\end{center}

In this section, we investigate the cavity-modified vibrational spectra of monolayer MoS$_2$. Pristine monolayer MoS$_2$ belongs to the point group D$_{3h}$, with its principal axis oriented along the $z$-direction, as illustrated in Fig.~\ref{fig:MoS2_IR}(a). 
The irreps of the vibrational modes $\Gamma^{\rm vib}$ at the $\Gamma$ point can be obtained from the direct sum decomposition of the direct product of the atomic-site representation and the polar vector representation~\cite{dresselhaus_group_theory}.
Among these modes, three correspond to the acoustic modes, 
while the remaining six are optical modes:
\begin{eqnarray}
\label{eq: mos2 optical modes}
    \Gamma^{\rm opt} =  A'_1({\rm R}) + E' ( {\rm IR+R} ) + A''_2 ( {\rm IR} ) + E''(\rm R) ,
\end{eqnarray}
where IR and R in the parentheses denote IR and Raman activity. The calculated IR and Raman spectra of pristine monolayer MoS$_2$ are shown as black solid lines in Figs.~\ref{fig:MoS2_IR}(c) and (d). The IR spectrum exhibits a peak at $\sim$ 373 cm$^{-1}$, corresponding to the $E'$ vibrational mode. The inset illustrates one of its eigenmodes, in which the Mo and S atoms vibrate in opposite directions along the $y$ axis. The Raman spectrum exhibits two peaks at $\sim $ 373 and 393 cm$^{-1}$, corresponding to the $E'$ and $A'_1$ modes, respectively. In this section, we focus only on these two modes since they have strong IR or Raman intensities. The remaining two optical modes in Eq.~(\ref{eq: mos2 optical modes}) are orders of magnitude weaker in intensity, as discussed below.

According to Tab.~\ref{tab: C and D group}, in the presence of a $y$-polarized (along the [120] direction) cavity mode, the symmetry of MoS$_2$ is reduced to C$_{2v}$, with the new principal axis aligned along $y$, as illustrated in Fig.~\ref{fig:MoS2_IR}(b). Because the principal axis is reoriented, the mirror plane $\mathcal{M}_z$ now contains the principal axis instead of being perpendicular to it. In standard references, the basis functions in a point-group character table are defined with the principal axis along $z$. For clarity and convenience, in Appendix~\ref{app: charater tables} we provide the character tables of D$_{3h}$ and C$_{2v}$ with the principal axes along $z$ and $y$, respectively. Under the symmetry reduction to C$_{2v}$, the irreps decompose as:
\begin{eqnarray}
\label{eq: A'1}
    A'_1(\rm R)  &\to& A_1 ({\rm IR+R}), \\
\label{eq: E'}
    E' ({\rm IR+R}) &\to& A_1 ({\rm IR+R}) + B_2 ({\rm IR+R}).
\end{eqnarray}
Accordingly, two main features are expected in the vibrational spectra. 
First, a new IR peak should emerge in the IR spectrum, since a previously IR-inactive mode becomes IR-active upon symmetry lowering, as shown in Eq.~(\ref{eq: A'1}). Second, the doubly degenerate mode $E'$ would split into two distinct peaks in both IR and Raman spectra, as shown in Eq.~(\ref{eq: E'}). 

To validate these predictions, we perform QEDFT calculations for monolayer MoS$_2$ coupled to a $y$-polarized cavity mode.
Consistent with Eq.~(\ref{eq: A'1}), a new IR peak with irrep $A_1$ emerges at $\sim 400$ cm$^{-1}$, as shown in Fig.~\ref{fig:MoS2_IR}(c).
Furthermore, the doubly degenerate IR peak splits into two, in agreement with Eq.~(\ref{eq: E'}).
In addition to peak splitting, there is also vibrational frequency renormalization due to vacuum fluctuations. The magnitudes of both effects are governed by the coupling ratio $\lambda/\omega$.
As shown in Fig.~\ref{fig:MoS2_IR}(c), the vibrational modes at the $\Gamma$ point in the presence of a $y$-polarized cavity mode closely resemble those of an H$_2$O molecule~\cite{dresselhaus_group_theory}, since both systems share the same $C_{2v}$ point group symmetry. The two $A_1$ modes in Fig.~\ref{fig:MoS2_IR}(c) can be viewed as linear combinations of breathing and symmetric stretching modes. Since $A_1$ is the totally symmetric representation, the associated vibrational modes preserve all symmetry operations of the group. In contrast, the $B_2$ mode transforms in the same way as $R_z$ (rotation about the $z$ axis), which is odd under $\mathcal C_2$ and $\mathcal{M}_x$ operations, consistent with Tab.~\ref{tab: C2v}.
The Raman spectrum in Fig.~\ref{fig:MoS2_IR}(d) likewise exhibits splitting of the $E'$ peak and frequency renormalization, in agreement with Eqs.~(\ref{eq: A'1}) and (\ref{eq: E'}).

It is important to note that symmetry analysis alone does not determine the intensity of a spectral peak. Whether a vibrational mode appears with appreciable intensity must be assessed through explicit scattering calculations or experimental measurements. For example, in monolayer MoS$_2$, both $A''_2$ and $E'$ vibrational modes are IR-active by symmetry. However, our density functional perturbation theory calculations show that the IR intensity of the $A''_2$ peak is about two orders of magnitude weaker than that of the $E'$ peak, as shown in Fig.~\ref{fig:MoS2_IR_higher_freq}. Likewise, the $E''$ mode is also Raman-active, but its Raman intensity is more than three orders of magnitude weaker than that of the peaks discussed in this section. As a result, although the doubly degenerate $E''$ Raman peak also splits under a $y$-polarized cavity mode, the intensities remain very weak, as shown in Fig.~\ref{fig:MoS2_Rm_higher_freq}.

\section{Discussion}

\label{sec: discussion}
\subsection{Cavity-induced macroscopic response}

In the previous section, we discussed cavity-induced modifications of microscopic properties, such as energy-level degeneracies and vibrational selection rules, arising from symmetry breaking. 
Beyond these microscopic effects, cavity light-matter coupling can also be used to manipulate macroscopic responses~\cite{libor_prx_altermagnet_2022,lin_spontaneous_2026, JKQ_prl}. Specifically, according to Tabs.~\ref{tab: C and D group} and~\ref{tab: T and O group}, a nonpolar PG may be reduced to a polar subgroup upon coupling to appropriate cavity modes. As a consequence, a material with zero net polarization can develop a finite polarization under cavity light-matter interaction. 
For instance, Tab.~\ref{tab: C and D group} shows that a cavity mode polarized along either $x$ or $y$, or two cavity modes polarized in the $(x+z)$ or $(y+z)$ plane, can reduce the nonpolar PG D$_3$ to its polar subgroup C$_2$, indicating that a cavity-induced polarization {along the $C_2$ axis} is symmetry-allowed. In the weak-coupling regime, we can consider the dressed cavity field as a perturbation, and the magnitude and direction of the induced polarization can be described by a leading-order third-rank response tensor~\cite{JKQ_prl}. 
The independent tensor elements are constrained by the PG symmetry of the material outside the cavity.

However, when the light-matter interaction is strong, higher-order contributions can become important, and the induced macroscopic response can deviate from the leading-order tensor description. In this regime, it is more appropriate to analyze the symmetry-broken subgroup of the cavity-coupled system directly. The subgroup-based analysis in this paper does not rely on a perturbative expansion in the cavity coupling strength.

\subsection{Long-wavelength limit and beyond}
A central assumption underlying our analysis above is the long-wavelength (dipole) approximation, in which the cavity field is effectively homogeneous on the scale of the unit cell and can be idealized as having D$_{\infty h}$ symmetry. In this regime, the cavity does not break translational invariance, and the inversion symmetry is also preserved. This approximation is suited to standard Fabry-P\'erot, split-ring, and surface cavities, and it also underlies earlier symmetry-based analyses of cavity-modified chemical reaction rates, such as the work of Jayachandran {\it et al}.~(Ref.~\cite{cavity_molecule_symmetry}) on vibrational strong coupling. 
Their work focuses on a specific cavity configuration: two orthogonal cavity modes polarized in $(x+y)$ coupled to uniaxial molecules whose principal axis is along $z$. 
In this setting, reaction-rate changes are rationalized via a phenomenological selection rule: the rate is enhanced when the tensor product of the vibrational and cavity-mode irreducible representations contains that of the reaction coordinate, i.e., $\Gamma^{\rm vib}\otimes\Gamma^{\rm cav} \supset \Gamma^{\rm rea}$. In our formalism, by contrast, the cavity influence is absorbed into the electronic degrees of freedom. We therefore do not explicitly assign an irrep to the cavity mode~\footnote{Actually, it is not always possible to assign cavity modes an irrep of the material's point group if the material or the cavity is allowed to rotate in 3D. But one can always analyze the symmetry reduction induced on the material's electron density}; instead, we characterize cavity effects through the symmetry reduction it induces on the electron density.

Looking forward, an important avenue is to relax the long-wavelength approximation and incorporate the finite spatial profile and wavelength of the cavity photon modes. Spatially structured cavity fields can break translational symmetry, introduce new length scales, and effectively generate moir\'e-like superstructures, thereby opening routes to cavity-controlled band folding and miniband engineering beyond what is possible with uniform fields \cite{Francesco_NC_moire_cavity}. On larger scales, the macroscopic cavity geometry itself---such as triangular Fabry-P\'erot mirrors with effective D$_{3h}$ symmetry or chiral cavities supporting a single circularly polarized mode~\cite{Hannes_chiral_cavity,chiral_cavity_exp_2022,xinle_cavity_axion,kono_chiral_cavity}---can 
even remove inversion or time-reversal symmetry, 
leading to new classes of cavity-induced phenomena that lie beyond a strictly unit-cell-based description. 
Thus, the present 32-point-group classification provides a building block toward a more general symmetry framework for a wider range of cavity configurations. The same group-theoretical strategy can be naturally extended to space groups, magnetic groups, spin groups, and related symmetry settings.
Extending the present group-theoretical analysis to these regimes, and combining it with real-space or multiscale electronic-structure methods, is a promising direction for future work.

\section{Conclusion}
\label{sec: conclusion}

In summary, our work establishes a unified group-theoretical framework for understanding how linearly polarized vacuum photon modes reshape the symmetry of crystalline materials within the long-wavelength approximation. By analyzing the symmetry transformation properties of the PF Hamiltonian, the effective pf QED interaction kernel \((\hat{\bf \Pi}\cdot\tilde{\boldsymbol{\epsilon}})^2\), and its pxLDA counterpart \((\mathbf{k}\cdot\tilde{\boldsymbol{\epsilon}})^2\), we classify the symmetry reductions induced by such modes 
and provide explicit subgroup relations for all 32 crystallographic point groups. 
This classification directly links the cavity configuration to material symmetry and thereby to various effects such as modifications in band structures and vibrational selection rules.

The power of this symmetry-first perspective is illustrated by several concrete applications. For cubic BaTiO$_3$, we show that cavity modes polarized in the (110), (111), and (001) planes reduce the point-group symmetry from O$_h$ to D$_{2h}$, D$_{3d}$, and D$_{4h}$, respectively. These distinct cavity-mode configurations therefore lead to different patterns of band-degeneracy lifting. In particular, the threefold and twofold degeneracies of the CB1 and CB2 at the $\Gamma$ point are either lifted or preserved, depending on the corresponding decomposition of the irreps in the symmetry-broken subgroups.
For monolayer MoS\(_2\), the reduction of the point-group symmetry ${\rm D}_{3h}\to {\rm C}_{2v}$ under a \(y\)-polarized mode explains both the splitting of the formerly doubly degenerate \(E^\prime(\rm IR+R)\) mode and the activation of an additional \(A_1\) IR peak.

More broadly, the symmetry principles developed here provide a blueprint for ``symmetry-guided'' cavity materials engineering. The same tools that predict electronic states and  vibrational peak splittings can be systematically applied to anticipate cavity-modified phonon dispersions, topological phase transitions, and selection rules in nonlinear and time-resolved spectroscopies. 
Furthermore, this framework can also be generalized to describe cavity-induced macroscopic responses like polarization and magnetization.
In this sense, our classification of symmetry breaking by dark, linearly polarized cavities provides a starting point for designing light-matter settings where desired material functionalities---electronic, vibrational, or magnetic---are targeted at the level of symmetry before extensive numerical or experimental effort is invested while simultaneously deepening our fundamental understanding. As cavity platforms, electronic-structure methods, and QEDFT functionals continue to advance, we expect such symmetry-based strategies to play a central role in shaping the next generation of quantum materials engineering in optical environments.

\section*{Acknowledgments}
We would like to thank Zhiyang Zeng, Xinle Cheng, Michael Ruggenthaler, and Hannes H{\"u}bener for fruitful discussions. We thank Professor Peizhe Tang for his critical reading of and valuable feedback on the manuscript. J.Q. acknowledges support from
the Max-Planck Graduate Center for Quantum Materials. We acknowledge support from the Cluster of Excellence “CUI: Advanced Imaging of Matter”-EXC 2056-project ID 390715994, SFB-925 “Light induced dynamics and control of correlated quantum systems”-project ID 170620586 of the Deutsche Forschungsgemeinschaft (DFG), the European Research Council (ERC-2024-SyG-UnMySt-101167294), and the Max Planck-New York City Center for Non-Equilibrium Quantum Phenomena. The Flatiron Institute is a division of the Simons Foundation.

\begin{appendix}

\section{Coordinate systems in the symmetry-broken subgroup tables}
\label{app: PG coordinate}

\begin{center}
    \begin{figure}[ht]
        \centering
        \includegraphics[width=0.8\linewidth]{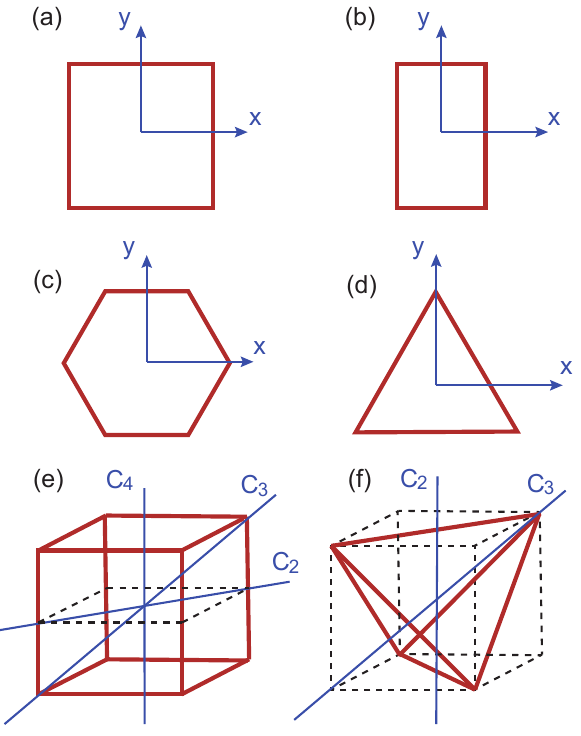}
        \caption{(a)-(d) Illustration of the coordinate system of the uniaxial PGs listed in Tab.~\ref{tab: C and D group}, with $\mathcal{C}_4$, $\mathcal{C}_2$, $\mathcal{C}_6$, and $\mathcal{C}_3$ rotation axes as the principal axes, respectively. (e) Illustration of the high symmetry axes in the $\rm O$ and ${\rm O}_h$ groups. (f) Illustration of the high symmetry axes in the $\rm T$, ${\rm T}_h$, and ${\rm T}_d$ groups.
        }
        \label{fig:PG coordinates}
    \end{figure}
\end{center}

\section{Computational Details}
\label{app: computational details}

The symmetry analysis in this paper makes use of the tools available on the Bilbao Crystallographic Server~\cite{bilbao_paper,bilbao_magnetic_group_tensor}. The DFT calculations are performed using the {\tt Quantum ESPRESSO} package~\cite{qe_2020}. The pxLDA functional is implemented in-house~\cite{I-te_pxLDA,benshu_qedft} within {\tt Quantum ESPRESSO}.
Cubic BaTiO$_3$ is calculated using optimized norm-conserving Vanderbilt pseudopotentials (ONCVPSPs)~\cite{ONCV_pseudopotential} version 0.5 from {\tt PseudoDojo}~\cite{pseudodojo};  monolayer MoS$_2$ is calculated using ONCVPSPs version 0.4.1 from {\tt PseudoDojo}; and graphene is calculated using an ultrasoft pseudopotential~\cite{ultrasoft} generated by {\tt PSlibrary}~\cite{PSlibrary}.
The Perdew-Burke-Ernzerhof exchange correlation functional~\cite{pbe} is employed for cubic BaTiO$_3$ and graphene, whereas the Perdew-Wang functional~\cite{PW_LDA} is employed for monolayer MoS$_2$. For BaTiO$_3$, a kinetic-energy cutoff of 84 Ry for wavefunctions, an $11\times11\times11$ $\bf k$-grid, and an energy convergence threshold of $10^{-10}$ Ry are employed for self-consistent calculations both with and without cavity photon modes. For MoS$_2$ and graphene, a kinetic-energy cutoff of 80 Ry for wavefunctions, a $16\times 16\times 1$ $\bf k$-grid, and an energy convergence threshold of $10^{-10}$ Ry  are employed. The corresponding lattice parameters are 4.008~\AA\ for cubic BaTiO$_3$, 3.159~\AA\ for monolayer MoS$_2$ with a vacuum spacing of 36.164~\AA, and 2.463~\AA\ for graphene with a vacuum spacing of 24.628~\AA.
In the band structure calculations of cubic BaTiO$_3$ with QEDFT, $\bm \epsilon_1=1/\sqrt{2}(1,-1,0), \bm \epsilon_2=(0,0,1)$,  $\bm \epsilon_1=1/\sqrt{2}(1,-1,0), \bm \epsilon_2=1/\sqrt{6}(1,1,-2)$, and $\bm \epsilon_1=(1,0,0), \bm \epsilon_2=(0,1,0)$ with $\lambda/\omega = 0.2$ are used to represent cavity modes in the (110), (111), and (001) planes, respectively.
For MoS$_2$ and graphene, ${\bm \epsilon} = (0,1,0)$ and $\lambda/\omega = 1.0$ are used as input values. Choosing other values of the coupling ratio $\lambda/\omega$ does not significantly affect any group-theoretical discussion in this work. 
To remain safely within the high-frequency regime where pxLDA is expected to be applicable, the cavity frequency $\omega$ should be sufficiently large compared with the relevant electronic excitation energies. In practice, a suitable value can be estimated from the material's dielectric function or optical conductivity: the cavity photon energy should lie well above the dominant optically active transitions relevant to the electronic states of interest. In this work, we therefore choose $\omega=10$ eV. However, we emphasize that our symmetry analysis does not rely on the high-frequency approximation as shown in Section~\ref{sec: qed hamiltonian}.
Note that the bare photon mode parameters are transformed into the dressed photon mode parameters $\tilde{\bm \epsilon}$ and $\tilde{\lambda}/\tilde{\omega}$ by a Bogoliubov transformation~\cite{I-te_pxLDA} within the code. 
For MoS$_2$, the dynamical matrix is further corrected by enforcing the translational acoustic sum rules~\cite{phonon_sum_rules}. The infrared and Raman intensities are calculated using density functional perturbation theory~\cite{QE_Raman}. For visualization, the resulting spectra are broadened using Gaussian functions with a linewidth of 3~${\rm cm}^{-1}$.

\section{Two cavity modes with unequal Coupling strengths}
\label{app: unequal coupling}

\begin{figure}[h]
    \centering
    \includegraphics[width=0.8\linewidth]{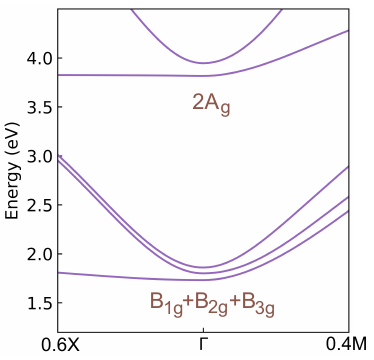}
    \caption{Band structure of BaTiO$_3$ coupled to two cavity photon modes with unequal coupling strengths, $\lambda/\omega=0.2$ and $\lambda/\omega=0.1$ along the $[100]$ and $[010]$ directions, respectively.}
    \label{fig:BTO x2y1}
\end{figure}

The methodology developed in this paper can be naturally extended to multiple modes with unequal coupling strengths. Here, we provide an example with unequal coupling strengths, $\lambda/\omega=0.2$ and $\lambda/\omega=0.1$, along the $[100]$ and $[010]$ directions of BaTiO$_3$, respectively. Under this cavity configuration, the three crystallographic directions of cubic BaTiO$_3$ become inequivalent, reducing the point-group symmetry from ${\rm O}_h$ to ${\rm D}_{2h}$. As shown in Fig.~\ref{fig:BTO x2y1}, the resulting lifting of the band degeneracies is fully consistent with the irreducible representation decomposition associated with ${\rm O}_h\rightarrow{\rm D}_{2h}$, as given in Eq.~(\ref{eq: Oh to D2h}).
More generally, the same symmetry-analysis framework can be straightforwardly extended to an arbitrary number of cavity photon modes with mode-dependent coupling strengths.

\section{Shift of the Dirac Cone in Graphene}
\label{app: Dirac cone shift}

\begin{center}
    \begin{figure}[h]
        \centering
        \includegraphics[width=1\linewidth]{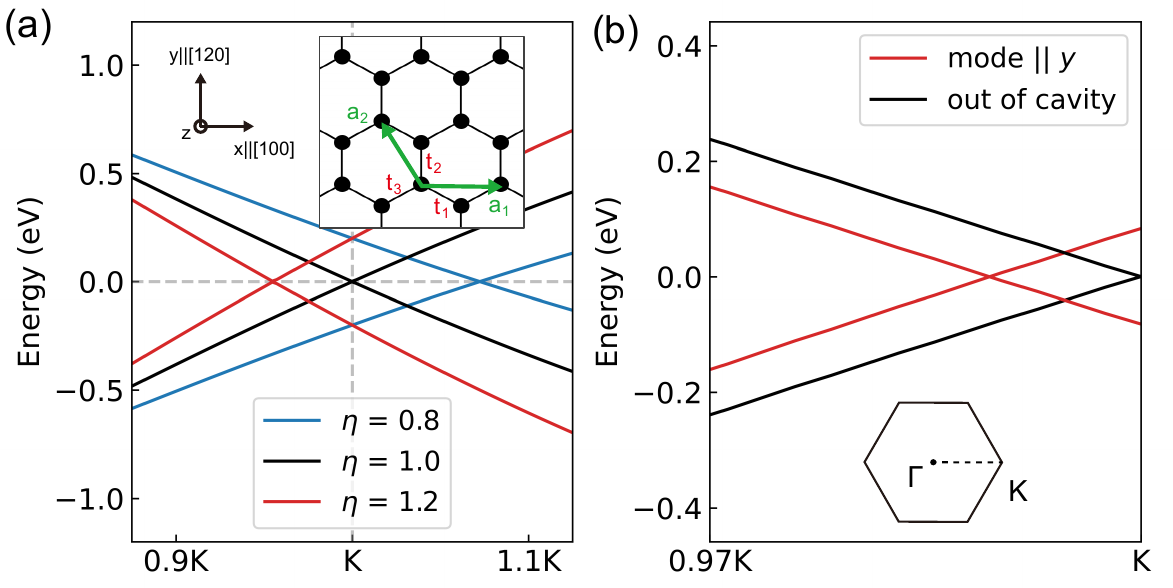}
        \caption{(a)
        Shift of the Dirac point in graphene calculated using a nearest-neighbor tight-binding model. The inset shows the primitive cell used in our calculations.
        (b) QEDFT band structure of graphene coupled to a $y$-polarized cavity mode near the $\rm K$ point.
        }
        \label{fig:graphene_band}
    \end{figure}
\end{center}

Graphene is one of the most prominent two-dimensional materials, characterized by Dirac cones centered at high-symmetry points $\rm K$ and $\rm K'$ in the Brillouin zone. 
Here we investigate how the Dirac cone of graphene is modified when it is coupled to a linearly polarized cavity mode. 
The little group at the $\rm K$ point is D$_{3h}$. According to Tab.~\ref{tab: C and D group}, a $y$-polarized mode lowers the group to C$_{2v}$ by breaking the $\mathcal C_3$ rotational axis.
Consequently, the Dirac cone is expected to shift away from the K or K$'$ point, which can be understood within a simple two-band tight-binding model~\cite{graphene_strain_prb}. 
As illustrated in the inset of Fig.~\ref{fig:graphene_band}(a), we adopt a nearest-neighbor tight-binding model with lattice vectors ${\bf a}_1 =(a, 0), {\bf a}_2 = (-a/2,\sqrt{3}a/2)$, and three hopping parameters $t_1$, $t_2$, and $t_3$.
In this model, the band dispersion of graphene is given by
\begin{equation}
    E({\bf k}) = \pm \left|e^{i{\bf k}\cdot \delta_2}t_2\left[1+2\eta e^{-i\frac{\sqrt{3}}{2}k_ya}\cos \left (\frac{1}{2}k_xa \right ) \right] \right| ,
\end{equation}
where $\delta_2$ denotes the C-C bond vector associated with hopping $t_2=1$, and $\eta = {t_1}/{t_2}={t_3}/{t_2}$ quantifies the anisotropy among the hopping terms. Along the $k_y = 0$ $\bf k$-path (i.e., $\Gamma$-$\rm{K}$), the Dirac point position is determined by the zeros of $E({\bf k})$.
In pristine graphene, the three nearest-neighbor hoppings are equivalent ($\eta = 1$) due to the $\mathcal{C}_3$ symmetry. Therefore, the Dirac point is located at $k_x = 4\pi /3 a$, corresponding to the high-symmetry point K. However, when a $y$-polarized cavity mode is introduced, the hopping parameter $t_2$ becomes inequivalent to $t_1$ and $t_3$, and thus $\eta \neq 1$. As a result, the zero of $E({\bf k})$ moves away from K due to the loss of rotational symmetry, as illustrated in Fig.~\ref{fig:graphene_band}(a). 

This prediction is confirmed by our QEDFT calculation, as shown in Fig.~\ref{fig:graphene_band}(b). Note that the Dirac cones in graphene are not protected by rotational symmetry but by $\mathcal{PT}$ symmetry~\cite{Dirac_cone_existence_prb}. Therefore, breaking $\mathcal C_3$ rotational symmetry can shift the Dirac points without opening a gap. A gap can emerge if the $\mathcal{PT}$ symmetry is broken, or if the perturbation becomes sufficiently strong to drive the two Dirac cones together and induce their annihilation. These situations, however, lie beyond the range accessible in our QEDFT framework.

\section{Character Tables}
\label{app: charater tables}

Here we provide the character table of the point group ${\rm D}_{3h}$ with the principal axis along $z$ in Tab.~\ref{tab: D3h}, and that of ${\rm C}_{2v}$ with the principal axis along $y$ in Tab.~\ref{tab: C2v}. Table~\ref{tab: C2v} is generated from the standard character table of ${\rm C}_{2v}$~\cite{bilbao_paper} by a cyclic permutation of the coordinate system: $(x,y,z)\to(z,x,y)$.

\begin{table}[h]
\centering
\begin{tabular}{c cccccc c c }
\toprule
D$_{3h}$ & $E$ & $\mathcal{M}_z$ & $2{\mathcal C}_3$ & $2{\mathcal S}_3$ & $3{\mathcal C}'_2$ & $3\mathcal{M}_v$ &  &  \\
\midrule
$A_1'$   & 1 & 1 & 1  & 1  & 1  & 1  &  & $x^2+y^2,\ z^2$ \\
$A_2'$   & 1 & 1 & $1$ & 1  & $-1$  & $-1$ & $R_z$ &  \\
$A''_1$     & 1 & $-1$ & 1  & $-1$  & $1$ & $-1$  &  \\
$A_2''$  & 1 & $-1$ & 1  & $-1$ & $-1$ & $1$ & $z$   \\
$E'$  & 2 & 2 & $-1$ & $-1$ & 0 & 0  & $(x,y)$ & $(x^2-y^2,\ xy)$ \\
$E''$    & 2 & $-2$ & $-1$  & 1 & 0  & 0  & $(R_x,R_y)$ & $(xz,\ yz)$ \\
\bottomrule
\end{tabular}
\caption{Character table of the D$_{3h}$ point group with the principal axis along the $z$ direction.}
\label{tab: D3h}
\end{table}

\begin{table}[h]
\centering
\begin{tabular}{c cccc c c }
\toprule
C$_{2v}$ & $E$ & ${\mathcal C}_2$ & $\mathcal{M}_x$ & $\mathcal{M}_z$ &  &  \\
\midrule
$A_1$   & 1 & 1 & 1  & 1  & $y$ & $x^2, \ y^2,\ z^2$ \\
$A_2$   & 1 & 1 & $-1$ & $-1$ & $R_y$ & $xz$ \\
$B_1$     & 1 & $-1$ & 1  & $-1$  & $R_x, \ z$ & $yz$  \\
$B_2$  & 1 & $-1$  & $-1$ & $1$ & $ R_z,\ x$ & $xy$   \\
\bottomrule
\end{tabular}
\caption{Character table of the C$_{2v}$ point group with the principal axis along the $y$ direction. Note that the standard character table is usually given with the principal axis along the $z$ direction, which will affect the definition of basis functions.}
\label{tab: C2v}
\end{table}

\section{Vibrational spectra in Other Frequency Ranges}
\label{app: spectrum smaller}
\begin{center}
    \begin{figure}[ht]
        \centering
        \includegraphics[width=0.9\linewidth]{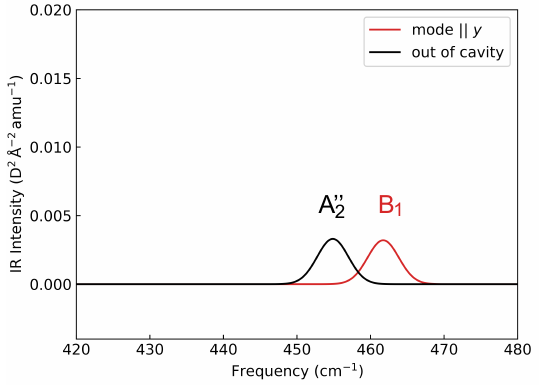}
        \caption{
        Infrared spectra of MoS$_2$ with and without the cavity mode in the frequency range 420-480 cm$^{-1}$, where the peak intensities are much smaller than those of the modes shown in the main text.
        }
        \label{fig:MoS2_IR_higher_freq}
    \end{figure}
\end{center}

\begin{center}
    \begin{figure}[ht]
        \centering
        \includegraphics[width=0.9\linewidth]{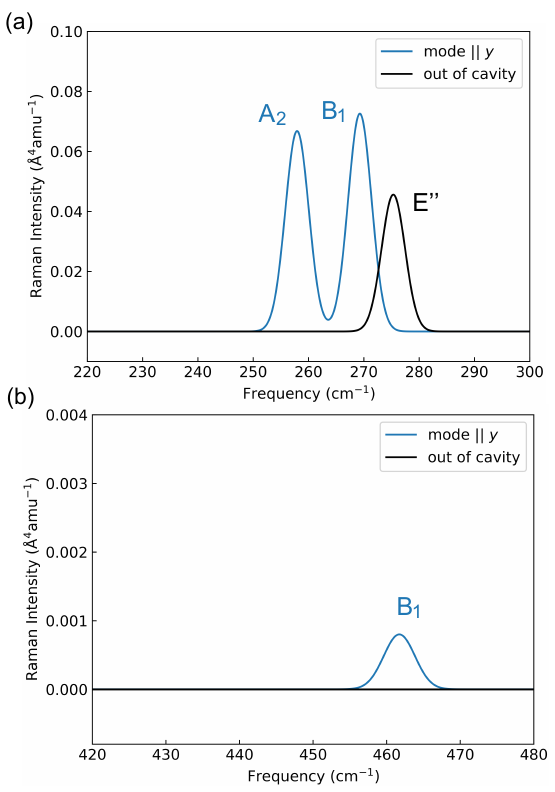}
        \caption{
        (a) Raman spectra of MoS$_2$ with and without the cavity mode in the frequency range 220-300 cm$^{-1}$, where the peak intensities are much smaller than those of the modes shown in the main text. (b) Same as (a), but in the frequency range 420-480 cm$^{-1}$.
        }
        \label{fig:MoS2_Rm_higher_freq}
    \end{figure}
\end{center}

\end{appendix}

\FloatBarrier


%

\end{document}